\documentclass[a4paper]{jpconf}
\usepackage{graphicx}

\def\lapp{\ifmmode\stackrel{<}{_{\sim}}\else$\stackrel{<}{_{\sim}}$\fi}
\def\gapp{\ifmmode\stackrel{>}{_{\sim}}\else$\stackrel{>}{_{\sim}}$\fi}
\def\degr{\ifmmode^{\circ}\else$^{\circ}$\fi}

\begin{document}
\title{Pulsar timing and its applications}

\author{R. N. Manchester}

\address{CSIRO Astronomy and Space Science, PO Box 76, Epping NSW 1710, Australia}

\ead{dick.manchester@csiro.au}

\begin{abstract}
  Pulsars are remarkably precise ``celestial clocks'' that can be used
  to explore many different aspects of physics and astrophysics. In
  this article I give a brief summary of pulsar properties and
  describe some of the applications of pulsar timing, including tests
  of theories of gravitation, efforts to detect low-frequency
  gravitational waves using pulsar timing arrays and establishment of
  a ``pulsar timescale''.
\end{abstract}

\section{Introduction}
Pulsars are remarkably precise ``celestial clocks''. They are believed
to be rotating neutron stars that send out beams of radio, optical,
X-ray and $\gamma$-ray emission which we observe as pulses when they
sweep across the Earth. About 2600 pulsars are currently known, almost
all of which lie within our Galaxy.\footnote{See the ATNF Pulsar
  Catalogue V1.56 (http://www.atnf.csiro.au/research/pulsar/psrcat)
  and \cite{mhth05}} The observed pulse periods range between 1~ms and
15~s, with most lying between 0.3~s and 3~s, the so-called ``normal
pulsars''. The ``millisecond pulsars'' (MSPs) form a distinct group,
most with periods in the range 1~ms to 10~ms. Although pulsar periods
($P$) are very stable and predictable, they are not constant. Most
pulsars are powered by their rotational kinetic energy and, as they
lose energy to relativistic particle winds and radiation, they slow
down. The rate of slowdown, $\dot P$, and other pulsar parameters can
be determined using pulse time-of-arrival (ToA) measurements made over
long intervals. MSPs are also distinguished by their very small $\dot
P$s, around five orders of magnitude smaller than $\dot P$s for normal
pulsars, and by the fact that most are in a binary orbit with another
star. These different properties result from the very different
evolutionary path followed by MSPs in which a central feature is
spin-up of an old neutron star by accretion from a companion star
\cite{bv91}.

In this review, we first discuss the basics of pulsar timing in
\S\ref{sec:timing}. Tests of theories of gravitation are discussed in
\S\ref{sec:grav}, and pulsar timing arrays (PTAs) and their use as
detectors of low-frequency gravitational radiation are described in
\S\ref{sec:pta}. Pulsar timescales and the application of pulsar
timing to navigation of distant spacecraft are discussed in
\S\ref{sec:timescale}. The main points of the review are briefly
summarised in \S\ref{sec:summary}.

\section{Pulsar timing}\label{sec:timing}
Measurement of a sequence of pulse ToAs over intervals ranging from
hours to decades is the basis of precision pulsar timing. These ToAs
are first transferred to an ``inertial'' reference frame, normally the
solar-system barycentre, to remove the effects of rotation and orbital
motion of the Earth. Radio-frequency dispersion resulting from the
presence of ionised gas in the interstellar medium can be compensated
for using observations in different frquency bands. Then, based on a
model for the intrinsic properties of the pulsar including its
astrometric, rotational, interstellar and (if applicable) binary
parameters, a predicted barycentric pulse arrival time is computed for
each ToA. The difference between the observed and predicted arrival
times is known as the timing ``residual''. Study of the time and
radio-frequency variations of these residuals is the basis of all
pulsar timing. Systematic variations of the residuals indicate either
errors in the pulsar model or the presence of unmodelled phenomena
affecting observed ToAs. Each pulsar parameter has a distinct
signature in the time and frequency domains. A least-squares fit of
the these signatures to the observed residuals enables corrections to
the various pulsar parameters and, possibly, the measurement of
previously un-measured parameters. Examination of the remaining
residuals enables investigation of new and different phenomena such as
relativistic binary perturbations or gravitational waves (GWs).

\begin{figure}[h]
\includegraphics[width=130mm]{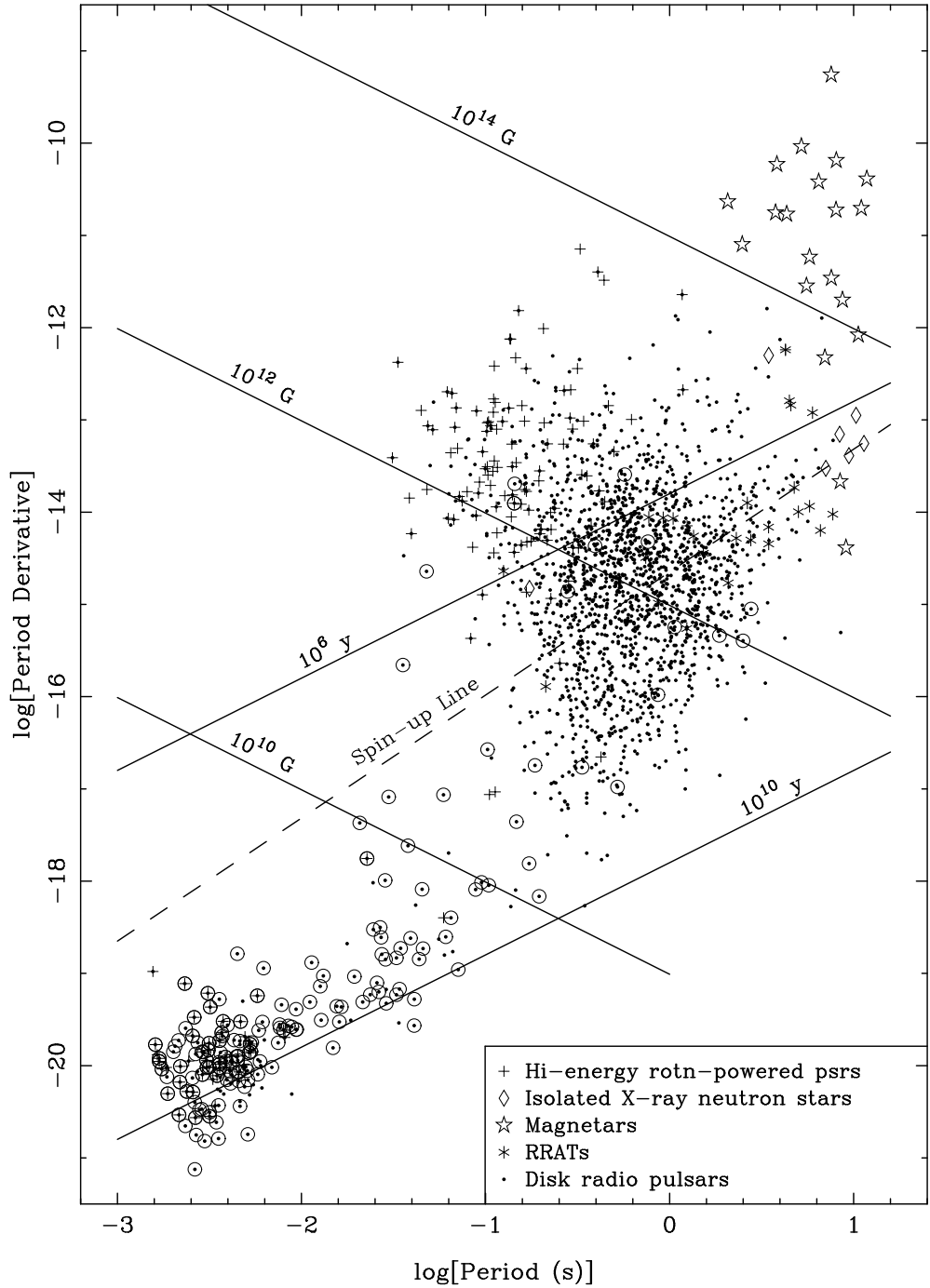}
\caption{Plot of rate of period increase, $\dot P$ versus pulse period
  $P$ for known pulsars lying in the Galactic disk, that is, excluding
  pulsars in globular clusters and extra-galactic pulsars. Different
  classes of pulsars are indicated by different symbols and binary
  pulsars are indicated by a circle around the pulsar symbol. Lines of
  constant characteristic age $\tau_c$ and surface dipole magnetic
  field strength $B_s$ are shown as is the maximimum spin period
  attainable by Eddington-limited accretion of mass from a companion
  (see, e.g., \cite{bv91}).  }\label{fg:ppdot}
\end{figure}

One of the most basic products of long-term pulsar timing is the rate
of slowdown of the pulsar rotation period $\dot P = - \dot\nu/\nu^2$,
where $\nu=1/P$ is the pulse frequency. Figure~\ref{fg:ppdot} shows
the plot of period derivative versus pulse period for the known
pulsars in the 
Galactic disk.\footnote{Globular cluster pulsars are omitted as their
  observed $\dot P$ is often affected by acceleration of the pulsar
  toward the cluster centre, see, e.g., \cite{frk+17}.} 
Assuming braking of the pulsar rotation by the reaction to emission of
magnetic-dipole radiation, that is, electromagnetic radiation at the
pulsar rotation frequency, and assuming a dipole field structure, we
can compute the magnetic field at the pulsar surface
\begin{equation}
  B_s = 3.2\times 10^{19} \sqrt{P \dot P},
\end{equation}
where $P$ is in seconds and $B_s$ is in Gauss,
and the ``characteristic age'', that is, the time since the
(back-extrapolated) pulse period was zero (see, e.g., \cite{mt77})
\begin{equation}
  \tau_c = P/(2\dot P).
\end{equation}

One of the main reasons why this diagram is so useful is that
different classes of pulsars generally lie in different zones on
it. For example, with their short period and low spin-down rate, MSPs
are located in the bottom left of the diagram. Conversely, because of
their typically long periods and strong magnetic fields, magnetars are
at the top right. Double-neutron-star (DNS) systems tend to lie in the gap
between MSPs and normal pulsars. Young pulsars have a high probability
of being detectable at high energies (optical and above) and the same
applies to MSPs having a larger-than-average spindown rate. The
so-called ``rotating radio transients'' (RRATs) (see, e.g.,
\cite{kkl+15}) tend to have long periods and to lie between the
magnetars and the bulk of normal pulsars. The X-ray isolated neutron
stars (XINS) (see, e.g., \cite{hab07}) also lie in this region.

The secular spin-down of pulsars can be described by the ``braking
index'', defined by
\begin{equation}
  n = \frac{\ddot\nu \nu}{\dot\nu^2}.
\end{equation}
Different values of $n$ are predicted for different braking
mechanisms. For example, $n=3$ for magnetic-dipole braking, $n=5$ for
braking by emission of GWs, and braking by pulsar
winds which deform the magnetosphere so that fields at the light
cylinder (where the co-rotation velocity equals the velocity of light)
are stronger than their dipole value have $1.0 \le n \le 3.0$
\cite{mt77}. Because of the combined effects of random
irregularities in pulsar period, especially in young pulsars, and the
relatively strong $\dot\nu$ dependence, braking indices representing
the secular braking of the pulsar have only been measured for a dozen
or so pulsars \cite{els17}. Most of these are between 2.0 and 3.0
indicating some braking by a stellar wind. For example, for the Crab
pulsar, the long-term average value of $n$ is $2.342\pm 0.001$
\cite{ljg+15}. 

\begin{figure}[h]
\includegraphics[width=130mm]{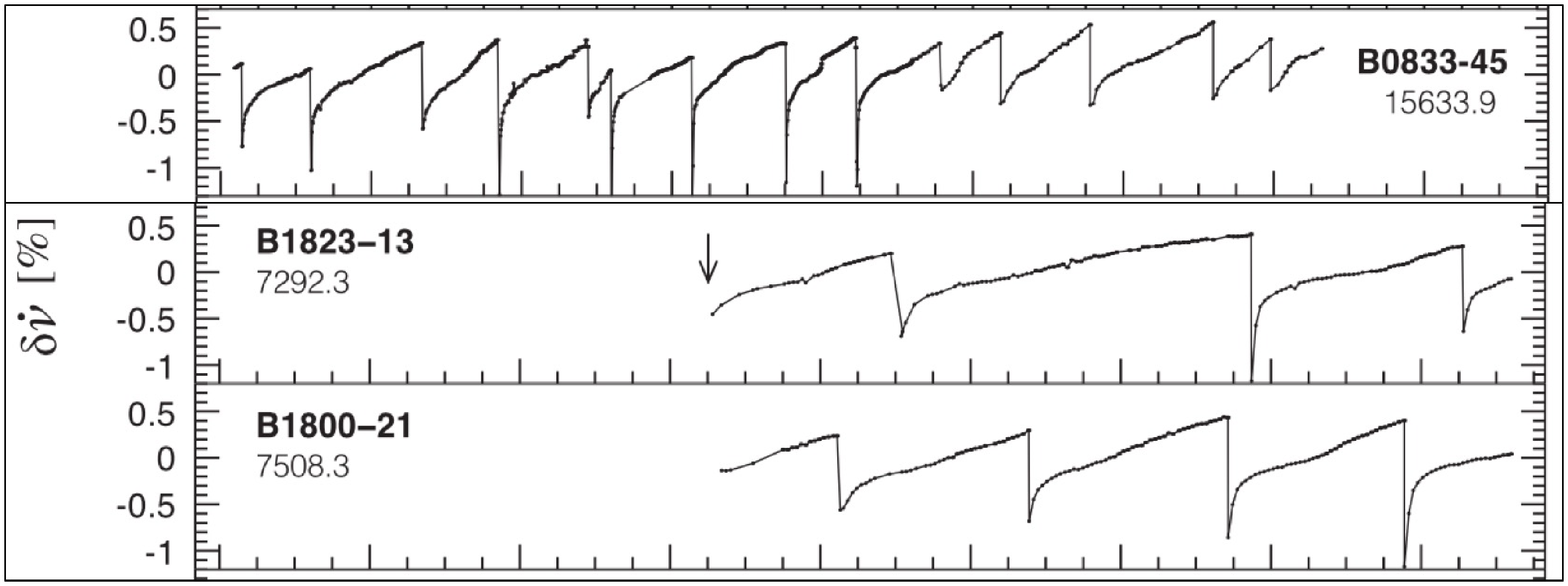}
\caption{Time dependence of spin-down rate $\dot\nu$ for three young
  pulsars that have multiple glitches. The horizontal axis is time
  over the Modified Julian Date range 40000 to 57500, a total of about
  48 years. The number below the pulsar name is the mean absolute
  spin-down rate in Hz s$^{-1}$. (Adapted from \cite{els17}).
}\label{fg:glitch}
\end{figure}

For young pulsars the regular slow-down is often interrupted by a
sudden spin-up known as a ``glitch''. About 425 of these glitches have
been observed in more than 140 pulsars\footnote{ATNF Pulsar Catalogue
  Glitch Table} and they have relative magnitudes 
$\Delta\nu_g/\nu$ of between $10^{-10}$ and $3\times 10^{-5}$ for
spin-powered pulsars. The post-glitch timing behaviour is different
for different pulsars, but for large glitches generally includes a
exponential relaxation toward the extrapolated pre-glitch
trend followed by an enhanced linear decay of $\dot\nu$. This
behaviour is illustrated in Figure~\ref{fg:glitch}.

During the linear decay phase, the apparent braking index is typically
between 20 and 40 \cite{els17}. However, as Figure~\ref{fg:glitch} shows, this
does represent the long-term evolution of $\dot\nu$. Following Lyne et
al. \cite{lpgc96}, Espinoza et al. \cite{els17} fitted a template to
the post-glitch changes in $\dot\nu$ to derive braking indices of
$1.7\pm 0.2$, $1.9\pm 0.5$ and $2.2\pm 0.6$ for PSRs B0833$-$45,
B1800$-$21 and B1823$-$13, respectively. These are all much less than
the magnetic-dipole value of 3.0 and less than other well-measured
secular braking indices. It should be noted, however, that Akbal et
al. \cite{aabp17} have independently derived a braking index for the
Vela pulsar (PSR B0833$-$45) of $2.81\pm 0.12$ by fitting to the
values of $\dot\nu$ just before a glitch rather than the post-glitch
variations fitted by \cite{els17}.

\section{Tests of gravitational theories}\label{sec:grav}
The discovery of the first binary pulsar, PSR B1913+16, by Hulse and
Taylor at Arecibo Observatory in 1974 \cite{ht75a} opened up multiple
new fields of pulsar research. Important among these was tests of
relativistic theories of gravitation. PSR B1913+16 was the first of
the class of DNS systems (now numbering 15) that have
short orbital periods, typically about a day, and relatively high
eccentricities. Together with the large system mass, these properties
imply large orbital velocities. For example, at periastron, the
orbital velocity $v$ of PSR B1913+16 and its companion are about 300
km~s$^{-1}$ or 0.1\% of the velocity of light. Since lowest-order
relativistic effects go as $(v/c)^2$, they are easily detectable as
orbital modulations of the observed pulsar period. Within a few years,
the first two relativistic perturbations, periastron advance and time
dilation, were detected \cite{tfm79}. In Einstein's general theory of
relativity (GR), these depend on the masses of the two stars, together
with the well-known Keplerian parameters. Consequently, the masses
could be determined and both were close to 1.4~M$_\odot$, confiming
that the system consisted of two neutron stars in orbit around one
another. Most importantly, measurement of the masses enabled
prediction of a third relativistic term, orbital decay due to the
emission of GWs from the system. This was also
measured in 1979 \cite{tfm79} and was in accordance with the
predictions of GR. Later measurements \cite{wnt10} show that the
observed orbit decay $\dot P_b$ is within 0.2\% of the GR
prediction. These observations are therefore, not only a confirmation
of the accuracy of GR as a theory of relativistic gravitation, but
also the first observational evidence for the existence of
GWs. These important results were recognised by the
award of the 1993 Nobel Prize in Physics to Joseph Taylor and Russell
Hulse.

An even more amazing binary system, the Double Pulsar, PSR
J0737$-$3039A/B, was discovered at Parkes in 2003
\cite{bdp+03,lbk+04}. This system has a orbital period of 2.4~h, about
one third that of PSR B1913+16 and is a DNS system. However, unlike
any other DNS system, both neutron stars are detectable as
pulsars. The A pulsar, the first-born neutron star of the two, was
spun up to its short period of 22~ms by accretion of mass and angular
momentum from its evolving companion which subsequently collapsed to
form the second neutron star. This is now detected as a relatively
young pulsar (B) with a pulse period of 2.77~s.\footnote{The B pulsar
  became undetectable in 2008, most likely because of spin-axis
  precession, but it should reappear sometime within the next decade
  or so \cite{pmk+10}.} With a predicted periastron advance of
$16.9\degr$~yr$^{-1}$, relativistic effects are even larger than for
PSR B1913+16, and hence more stringent tests can be performed
\cite{ksm+06}. Six relativistic effects have now been detected for
this system, including two describing the Shapiro delay, geodetic
precession of the spin axis of pulsar B \cite{bkk+08}, and a much more
precise measurement of the orbital decay (Kramer et al., in
preparation). These results show that GR accurately describes the
system at the 0.02\% level, an order of magnitude better than the PSR
B1913+16 test and the most stringent test so far of GR in the
strong-field regime.

Despite the success of GR in accounting for the details of ToA orbital
modulations in the DNS systems, it remains possible that departures
from the predictions of GR will be found in the future. There are many
alternative theories of gravity and pulsars provide several mechanisms
for testing the parameters of such theories. Some of these theories
violate the Strong Equivalence Principle (SEP), for example, allowing
differences between the gravitational and inertial masses. As pointed
out by \cite{ds91}, this would result in bodies with different
gravitational self-energy falling at different rates in an external
gravitational field. For binary pulsars with white dwarf or other
low-mass companions, an orbital eccentricity would be induced by the
gravitational field of the Galaxy. A limit on this, expressed in terms
of the parameter $|\Delta| = |m_g/m -1|$ of $\sim 4.6\times 10^{-3}$,
where $m_g$ and $m$ are the gravitational and interial masses respectively,
effectively of the neutron star, has been obtained from an analysis of
the orbital eccentricities of 27 pulsar -- white dwarf systems with
low intrinsic eccentricity \cite{gsf+11}.

Other gravitational theories predict the existence of dipolar
gravitational radiation. A strong limit on this has been placed by
Freire et al. \cite{fwe+12} from an analysis of the various
contributions to $\dot P_b$ in the pulsar -- white dwarf system PSR
J1738+0333.

The recent discovery of the very interesting triple system, PSR
J0337+1715, by Ransom et al. \cite{rsa+14} has opened up the
possibility of an even more stringent test of the SEP. This system
consists of a 2.73-ms pulsar in a 1.7-day orbit around a
0.19~M$_\odot$ white dwarf. In a co-planar orbit of period 327 days
around both these stars is a somewhat more massive 0.41~M$_\odot$
white dwarf. Aside from interesting questions about the orbital
dynamics and origin and evolution of this system, it is possible that
the gravitational field of the outer white dwarf will induce an
eccentricity in the orbit of the inner white dwarf. Since the
gravitational field of the outer white dwarf at the position of the
inner white dwarf is about $10^6$ times as strong as the Galactic
gravitational field, this test is potentially much more sensitive than
the SEP test based on the pulsar -- white-dwarf binary systems
\cite{gsf+11}. 

\section{Pulsar timing arrays}\label{sec:pta}
Although MSPs have extremely stable periods, one cannot rule out the
possibilty of intrinsic variations in the observed timing parameters
of a given pulsar. Uncorrected variations in interstellar propagation
delays are also possible. Therefore, in order to search for extrinsic
modulations due to (for example) low-frequency GWs propagating through
the Galaxy, it is necessary to observe many MSPs and to search for
correlated timing variations among the sample. Such a set of MSPs,
widely distributed on the sky and with frequent high-precision timing
observations over a long data span is known as a ``pulsar timing
array'' (PTA). PTAs are sensitive to low-frequency (nanoHertz)
gravitational waves that could be generated by orbiting super-massive
black holes in the cores of distant galaxies. Although individual
binary systems could be detectable, the most likely GW source to be
detected by PTAs is a stochastic background generated by many such
binary systems in galaxies at redshifts of between one and two
\cite{svc08}. Predicted amplitudes of such a background suggest that
observations of about 20 MSPs over 10 years are necessary for a
detection.

There are three main PTA projects around the world: the European Pulsar
Timing Array (EPTA) which uses data from several large radio
telescopes in Europe,  the North American NANOGrav array, which uses
data from Arecibo and the Green Bank Telescope, and the
Parkes Pulsar Timing Array (PPTA) which uses data from the Parkes 64-m
telescope in Australia. Each of these has high-precision timing data
on at least 20 MSPs with data spans ranging between a few years and
nearly 30 years. Typically the pulsars in each PTA are observed at
intervals of 2 -- 3 weeks in two or three different
radio-frequency bands.

Detection of GWs by PTAs relies on the correlations between the timing
signals for pulsars in different directions on the sky induced by GWs
passing over the Earth. Hellings \& Downs \cite{hd83} showed that, for
a stochastic background, the correlations depend only on the angular
separation of the pulsar pairs, not their sky location. Pulsars close
together on the sky have positively correlated signals but, because of
the quadrupolar nature of gravitational radiation, pulsars that are in
perpendicular directions have negatively correlated signals. The
correlation returns to a positive value for pulsars that are
diametrically opposed on the sky. GWs passing over the pulsars produce
signals that are uncorrelated between the pulsars in the array, just
adding a noise component to the GW signal.

Other correlated signals may exist in the data. For example,
irregularities in the reference timescale (typically International
Atomic Time, TAI), will have the same effect on timing residuals for
all pulsars in the array. In terms of the sky distribution, this is a
monopolar effect. Errors in the solar-system ephemeris used to refer
the ToAs to the solar-system barycentre are equivalent to errors in
the computed position of the Earth and so will have a dipolar
signature on the sky. These different dependencies on sky position can
be used to separate the different effects. 

\begin{figure}[h]
\includegraphics[width=130mm]{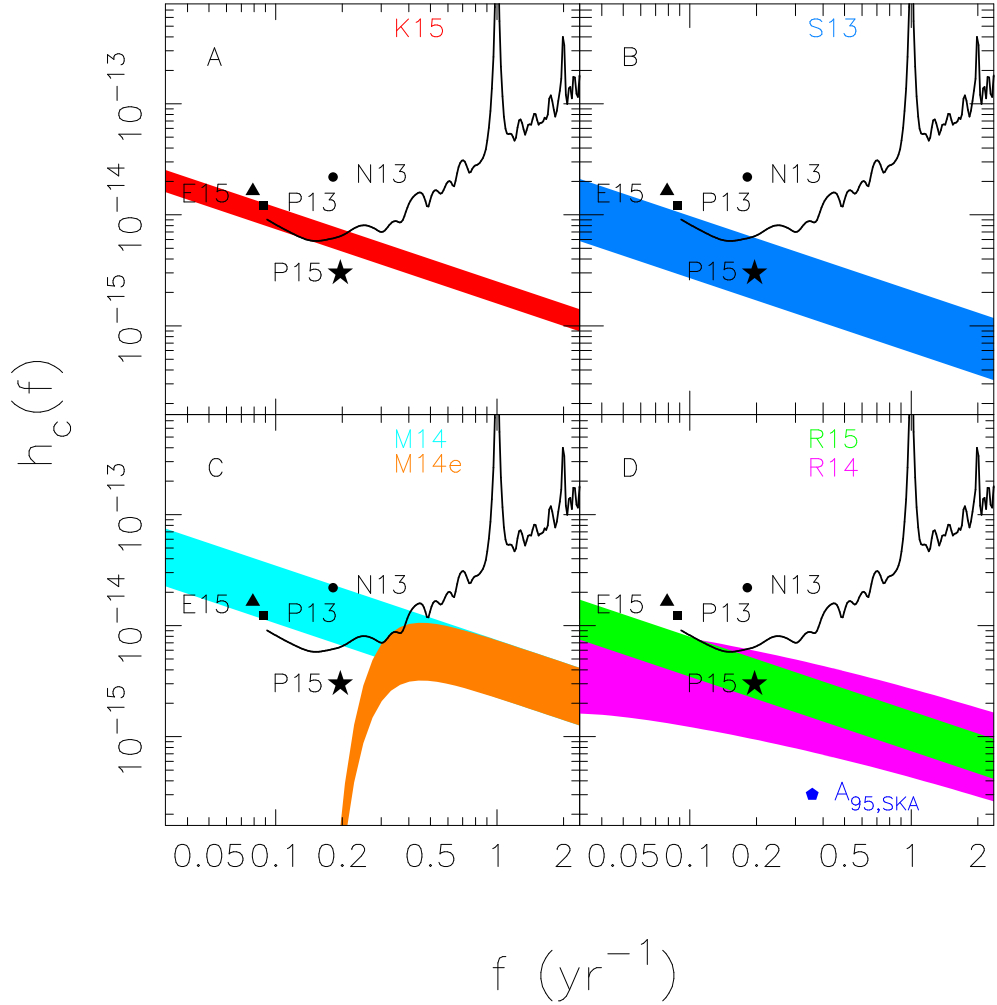}
\caption{Limits on the characteristic strain of a GW background in the
  Galaxy from PTA observations. Marked points are 95\% confidence
  limits from NANOGrav (N13, \cite{dfg+13}), the EPTA (E15,
  \cite{gs15}, the 2013 PPTA limit (P13, \cite{src+13}), and the most
  recent PPTA limit (P15, \cite{srl+15}). Different predictions for
  the spectrum of the GW background are shown in the different
  panels. The black line is the nominal spectral sensitivity of the
  PPTA observations used to derive the P15 limit and the point marked
  A$_{\rm 95,SKA}$ gives the expected sensitivity of a PTA based on
  the Square Kilometre Array. (From Shannon et al., Science, 2015,
  \cite{srl+15}. Reprinted with permission from
  AAAS.)}\label{fg:gw_limit}
\end{figure}

Up to now, there has been no positive detection of nanoHertz GWs by
pulsar timing arrays. However, just by using the best few pulsars in
an array, it is possible to set a limit on the strength of the GW
background in the Galaxy. The best limit so far is from the PPTA as
illustrated in Figure~\ref{fg:gw_limit}. This P15 limit, based on
Parkes observations in the 10cm band ($\sim 3$~GHz) of the four
best-performing MSPs in the PPTA sample, effectively rules out all
``standard'' models for generation of a low-frequency GW
background. This means that one or more of the assumptions that went
into the construction of these models, for example, the merger rate of
galaxies or the mass function for super-massive black holes in the
cores of galaxies, is in error. Shannon et al. \cite{srl+15} consider
that the most likely cause for the lower-than-expected GW background
level is that, at a late stage in their evolution, super-massive
black-hole binary systems lose energy to surrounding gas rather than
to GWs. It should be remarked that recently uncovered differences in
solar-system ephemerides may affect the derived GW background
limits.

An analysis of the GW senstivity of a PTA as a function of the PTA
parameters \cite{sejr13} shows that, when the self-noise from GWs
passing over the pulsars of a PTA is significant, the most efficient
way to increase the sensitivity of the array is to increase the number
of pulsars in it. To this end, the three regional PTAs have combined
to form the International Pulsar Timing Array (IPTA). Combined IPTA
data sets for 49 pulsars are now available \cite{vlh+16} and are
starting to be used to further PTA science objectives, e.g.,
\cite{lsc+16}. Other ways in which the number of pulsars in PTAs can
be increased is through finding previously unknown MSPs with good
timing properties in pulsar searches, e.g. \cite{sab+16}, and through
the use of new large radio telescopes such as the just-commissioned
Five hundred metre Aperture Spherical Telescope (FAST) in China
\cite{nlj+11} and the Square Kilometer Array (SKA), Phase 1 of which
should be operational around 2023 \cite{cr04b}.

\section{Pulsar timescales and navigation}\label{sec:timescale}
As mentioned in the previous section, PTAs are sensitive to
irregularities in the timescale to which the observations are
referenced, normally either TAI or the post-corrected version of TAI, BIPMxx,
where xx represents the year of creation \cite{pet03}. One can invert
the normal process of pulsar timing and use the pulsars to ``time the
timescale'', in effect establishing a ``pulsar timescale''. Such a
pulsar timescale differs from terrestrial timescales based on atomic
frequency standards in a number of important ways:
\begin{itemize}
\item it is independent of terrestrial timescales
\item it is based on entirely different physics -- the rotation of
    macroscopic bodies rather than quantum processes
\item it will be continuous for billions of years. 
\end{itemize}
Because of the nature of pulsar timing, pulsar timescales are only
competitive to (or of greater stability than) atomic timescales over
long intervals, months to years and decades. With their very different
basis, pulsar timescales can provide a valuable independent check on
the long-term stability of terrestrial atomic timescales \cite{hcm+12}.

\begin{figure}[h]
\includegraphics[width=130mm]{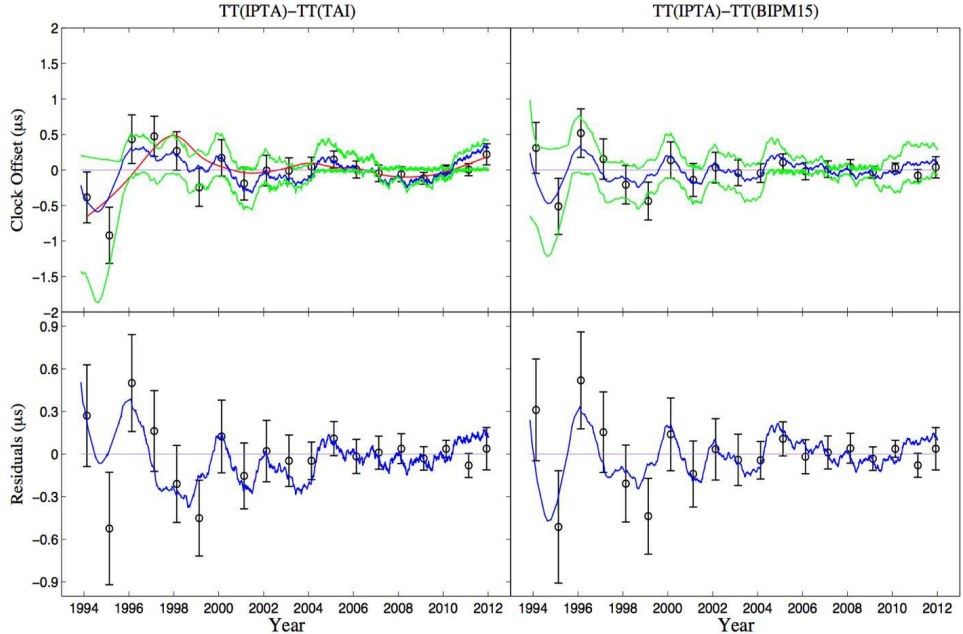}
\caption{Offsets of the pulsar timescale TT(IPTA16) with respect to
  TT(TAI) (left panels) and TT(BIPM15) (right panels) derived from a
  frequentist analysis (black circles with error bars) and a Bayesian
  analysis (blue line with 1-$\sigma$ error ranges in green) of the
  IPTA DR1 dataset. For the TT(TAI) reference, the red line gives the
  offset of TT(BIPM15) from TT(TAI) after subtracting quadratic
  terms. The lower panels show the residuals from TT(BIPM15). (Guo et
  al., in preparation)}\label{fg:timescale}
\end{figure}

Figure~\ref{fg:timescale} shows the results of analyses of the IPTA
DR1 dataset \cite{vlh+16} for the ``common-mode'' or clock term
relative to TT(TAI) and to TT(BIPM15)\footnote{Terrestrial Time (TT)
  is a uniform theoretical timescale based on the value of the SI
  second on the Earth's geoid. TT(TAI) and TT(BIPM15) are realisations
  of TT based on TAI and BIPM15 respectively.} using a frequentist
({\sc tempo2}) method and a Bayesian method (Guo et al., in
preparation). The pulsar offsets define a timescale which we label
TT(IPTA16). The left panels of Figure~\ref{fg:timescale} show that
TT(IPTA16) has significant deviations from TT(TAI) and, within the
uncertainties, is consistent with TT(BIPM15). Directly referencing the
pulsar ToAs to TT(BIPM15), as shown in the right panels, confirms the
agreement of the two timescales. The figure also shows that consistent
results are obtained with the two analysis methods. These results
demonstrate that the long-term stability of a timescale based on the
best available pulsar datasets is comparable to that of the best
available atomic timescales. Furthermore, we have shown that the
post-corrected timescale TT(BIPM15) is indeed more stable than
TT(TAI).

Another interesting application of precision pulsar timing is to
navigation of spacecraft that are distant from the Earth, even outside
the solar system. Pulsars can form a ``celestial GPS'' system --
instead of assuming a fixed observatory position and solving for the
pulsar parameters as in normal pulsar timing, one can analyse the ToAs
from a set of pulsars with known parameters and solve for the
observatory position. For spacecraft navigation, it is most practical
to use an X-ray telescope on the spacecraft and to observe a sample of
pulsars with significant X-ray pulsed emission. Analysis of a
realistic simulation \cite{dhy+13} has shown that position location
with an accuracy of better than 20~km is possible using observations
of just four MSPs. Autonomous operation of the system is possible, but
accuracy is improved with updates of the pulsar parameters from
Earth-based observations. It is interesting to note that in 2016 the
Chinese launched a satellite, XPNAV, dedicated to exploring X-ray
pulsar navigation, and that NASA's recently launched NICER mission on
the International Space Station has a project (SEXTANT) devoted to
this topic as well.

\section{Summary}\label{sec:summary}
Pulsar timing has many possible applications in physics and
astrophysics. Their remarkable intrinsic period stability, especially
for MSPs, their distribution throughout the Galaxy and the fact that
many pulsars, including more than half of all MSPs, are in binary
orbits around another star, allow explorations of theories of
gravitation, properties of the interstellar medium, studies of binary
and stellar evolution and many other topics. Pulsar timing arrays
(PTAs) can be used as detectors of low-frequency gravitational
radiation and to establish a ``pulsar timescale'' that is
independent of terrestrial atomic timescales. In most of these areas,
pulsars provide unique results and insights into the relevant physical
processes. With new and more sensitive radio telescopes coming soon,
the sample of pulsars that can contribute to these studies will grow,
giving higher precision for existing applications and opening up new
fields of research.

\ack I thank my colleagues for their many contributions to the topics
discussed and the referee for helpful suggestions.

\section*{References}


\begin{thebibliography}{10}
\expandafter\ifx\csname url\endcsname\relax
  \def\url#1{{\tt #1}}\fi
\expandafter\ifx\csname urlprefix\endcsname\relax\def\urlprefix{URL }\fi
\providecommand{\eprint}[2][]{\url{#2}}

\bibitem{mhth05}
{Manchester} R~N, {Hobbs} G~B, {Teoh} A and {Hobbs} M 2005 {\em AJ\/} {\bf 129}
  1993--2006

\bibitem{bv91}
Bhattacharya D and {van den Heuvel} E~P~J 1991 {\em Phys. Rep.\/} {\bf 203}
  1--124

\bibitem{frk+17}
{Freire} P~C~C, {Ridolfi} A, {Kramer} M, {Jordan} C, {Manchester} R~N, {Torne}
  P, {Sarkissian} J, {Heinke} C~O, {D'Amico} N, {Camilo} F, {Lorimer} D~R and
  {Lyne} A~G 2017 {\em MNRAS\/} {\bf 471} 857--876 

\bibitem{mt77}
Manchester R~N and Taylor J~H 1977 {\em Pulsars\/} (San Francisco: Freeman)

\bibitem{kkl+15}
{Karako-Argaman} C, {Kaspi} V~M, {Lynch} R~S, {Hessels} J~W~T, {Kondratiev}
  V~I, {McLaughlin} M~A, {Ransom} S~M, {Archibald} A~M, et al.
  2015 {\em ApJ\/} {\bf 809} 67 

\bibitem{hab07}
{Haberl} F 2007 {\em Astrophys. Space Sci.\/} {\bf 308} 181--190
  

\bibitem{els17}
{Espinoza} C~M, {Lyne} A~G and {Stappers} B~W 2017 {\em MNRAS\/} {\bf 466}
  147--162 

\bibitem{ljg+15}
{Lyne} A~G, {Jordan} C~A, {Graham-Smith} F, {Espinoza} C~M, {Stappers} B~W and
  {Weltevrede} P 2015 {\em MNRAS\/} {\bf 446} 857--864 

\bibitem{lpgc96}
Lyne A~G, Pritchard R~S, Graham-Smith F and Camilo F 1996 {\em Nature\/} {\bf
  381} 497--498

\bibitem{aabp17}
{Akbal} O, {Alpar} M~A, {Buchner} S and {Pines} D 2017 {\em MNRAS\/} {\bf 469}
  4183--4192 

\bibitem{ht75a}
Hulse R~A and Taylor J~H 1975 {\em ApJ\/} {\bf 195} L51--L53

\bibitem{tfm79}
Taylor J~H, Fowler L~A and McCulloch P~M 1979 {\em Nature\/} {\bf 277} 437

\bibitem{wnt10}
{Weisberg} J~M, {Nice} D~J and {Taylor} J~H 2010 {\em ApJ\/} {\bf 722}
  1030--1034 

\bibitem{bdp+03}
{Burgay} M, {D'Amico} N, {Possenti} A, {Manchester} R~N, {Lyne} A~G, {Joshi}
  B~C, {McLaughlin} M~A, {Kramer} M, {Sarkissian} J~M, {Camilo} F, {Kalogera}
  V, {Kim} C and {Lorimer} D~R 2003 {\em Nature\/} {\bf 426} 531--533

\bibitem{lbk+04}
Lyne A~G, Burgay M, Kramer M, Possenti A, Manchester R~N, Camilo F, McLaughlin
  M~A, Lorimer D~R, D'Amico N, Joshi B~C, Reynolds J and Freire P~C~C 2004 {\em
  Science\/} {\bf 303} 1153--1157

\bibitem{pmk+10}
{Perera} B~B~P, {McLaughlin} M~A, {Kramer} M, {Stairs} I~H, {Ferdman} R~D,
  {Freire} P~C~C, {Possenti} A, {Breton} R~P, {Manchester} R~N, {Burgay} M,
  {Lyne} A~G and {Camilo} F 2010 {\em ApJ\/} {\bf 721} 1193--1205
  
\bibitem{ksm+06}
{Kramer} M, {Stairs} I~H, {Manchester} R~N, {McLaughlin} M~A, {Lyne} A~G,
  {Ferdman} R~D, {Burgay} M, {Lorimer} D~R, {Possenti} A, {D'Amico} N,
  {Sarkissian} J~M, {Hobbs} G~B, {Reynolds} J~E, {Freire} P~C~C and {Camilo} F
  2006 {\em Science\/} {\bf 314} 97--102

\bibitem{bkk+08}
{Breton} R~P, {Kaspi} V~M, {Kramer} M, {McLaughlin} M~A, {Lyutikov} M, {Ransom}
  S~M, {Stairs} I~H, {Ferdman} R~D, {Camilo} F and {Possenti} A 2008 {\em
  Science\/} {\bf 321} 104-- 

\bibitem{ds91}
Damour T and Sch\"afer G 1991 {\em Phys. Rev. Lett.\/} {\bf 66} 2549

\bibitem{gsf+11}
{Gonzalez} M~E, {Stairs} I~H, {Ferdman} R~D, {Freire} P~C~C, {Nice} D~J,
  {Demorest} P~B, {Ransom} S~M, {Kramer} M, {Camilo} F, {Hobbs} G, {Manchester}
  R~N and {Lyne} A~G 2011 {\em ApJ\/} {\bf 743} 102 
  \eprint{1109.5638})

\bibitem{fwe+12}
{Freire} P~C~C, {Wex} N, {Esposito-Far{\`e}se} G, {Verbiest} J~P~W, {Bailes} M,
  {Jacoby} B~A, {Kramer} M, {Stairs} I~H, {Antoniadis} J and {Janssen} G~H 2012
  {\em MNRAS\/} {\bf 423} 3328--3343 

\bibitem{rsa+14}
{Ransom} S~M, {Stairs} I~H, {Archibald} A~M, {Hessels} J~W~T, {Kaplan} D~L,
  {van Kerkwijk} M~H, {Boyles} J, {Deller} A~T, et al. 2014 {\em Nature\/}
  {\bf 505} 520--524 

\bibitem{svc08}
{Sesana} A, {Vecchio} A and {Colacino} C~N 2008 {\em MNRAS\/} {\bf 390}
  192--209 

\bibitem{hd83}
Hellings R~W and Downs G~S 1983 {\em ApJ\/} {\bf 265} L39

\bibitem{srl+15}
{Shannon} R~M, {Ravi} V, {Lentati} L~T, {Lasky} P~D, {Hobbs} G, {Kerr} M,
  {Manchester} R~N, {Coles} W~A, et al.
  2015 {\em Science\/} {\bf 349} 1522--1525 

\bibitem{dfg+13} {Demorest} P~B, {Ferdman} R~D, {Gonzalez} M~E, {Nice}
  D, {Ransom} S, {Stairs} I~H, {Arzoumanian} Z, {Brazier} A, et
  al. 2013 {\em ApJ\/} {\bf 762} 94 

\bibitem{gs15}
{Gerosa} D and {Sesana} A 2015 {\em MNRAS\/} {\bf 446} 38--55
  

\bibitem{src+13}
{Shannon} R~M, {Ravi} V, {Coles} W~A, {Hobbs} G, {Keith} M~J, {Manchester} R~N,
  {Wyithe} J~S~B, {Bailes} M, et al. 2013 {\em Science\/} {\bf 342} 334--337 

\bibitem{sejr13}
{Siemens} X, {Ellis} J, {Jenet} F and {Romano} J~D 2013 {\em Class. Quant.
  Grav.\/} {\bf 30} 224015 

\bibitem{vlh+16}
{Verbiest} J~P~W, {Lentati} L, {Hobbs} G, {van Haasteren} R, {Demorest} P~B,
  {Janssen} G~H, {Wang} J~B, {Desvignes} G, et al. 2016 {\em MNRAS\/} {\bf 458} 1267--1288
  

\bibitem{lsc+16}
{Lentati} L, {Shannon} R~M, {Coles} W~A, {Verbiest} J~P~W, {van Haasteren} R,
  {Ellis} J~A, {Caballero} R~N, {Manchester} R~N, et al. 2016 {\em MNRAS\/} {\bf 458} 2161--2187
  

\bibitem{sab+16}
{Stovall} K, {Allen} B, {Bogdanov} S, {Brazier} A, {Camilo} F, {Cardoso} F,
  {Chatterjee} S, {Cordes} J~M, et al. 2016 {\em ApJ\/} {\bf 833} 192
  

\bibitem{nlj+11}
{Nan} R, {Li} D, {Jin} C, {Wang} Q, {Zhu} L, {Zhu} W, {Zhang} H, {Yue} Y and
  {Qian} L 2011 {\em Intnl J. of Mod. Phys. D\/} {\bf 20} 989--1024
  

\bibitem{cr04b}
Carilli C and Rawlings S 2004 {\em New Astronomy Reviews}, {\bf 48} 11--12

\bibitem{pet03}
{Petrova} S~A 2003 {\em A\&A\/} {\bf 408} 1057--1063

\bibitem{hcm+12}
{Hobbs} G, {Coles} W, {Manchester} R~N, {Keith} M~J, {Shannon} R~M, {Chen} D,
  {Bailes} M, {Bhat} N~D~R, et al. 2012 {\em MNRAS\/} {\bf 427} 2780--2787
  

\bibitem{dhy+13}
{Deng} X~P, {Hobbs} G, {You} X~P, {Li} M~T, {Keith} M~J, {Shannon} R~M, {Coles}
  W, {Manchester} R~N, {Zheng} J~H, et al. 2013
  {\em Adv. Space Res.\/} {\bf 52} 1602--1621 

\end{thebibliography}

\providecommand{\newblock}{}

\end{document}